\documentclass{mn2e}
\usepackage{epsfig}

\newif\ifAMStwofonts
\AMStwofontstrue

\title[Weakly  non-adiabatic  model]{A weakly  non-adiabatic  one-zone
       model of stellar pulsations:  application to Mira stars}

\author[A.~Munteanu,  E.~Garc\'{\i}a--Berro  \& J.~Jos\'e]
       {Andreea Munteanu$^{1}$, 
        Enrique Garc\'{\i}a--Berro$^{1,2}$, \&
        Jordi Jos\'{e}$^{2,3}$\\   
$^1$ Departament de F\'{\i}sica Aplicada, Universitat Polit\`ecnica de
     Catalunya,  Jordi Girona  Salgado s/n, M\`odul B--5, Campus Nord,
     \\ 08034 Barcelona, Spain \\
$^2$ Institut  d'Estudis  Espacials de Catalunya,  Edifici Nexus, Gran
     Capit\`a 2-4, 08034 Barcelona, Spain \\
$^3$ Departament  de  F\'{\i}sica  i Enginyeria  Nuclear,  Universitat
     Polit\`ecnica  de  Catalunya,  Av.  V\'\i ctor  Balaguer  s/n, \\
     08800 Vilanova i la Geltr\'u (Barcelona), Spain}

\begin{document}
\date{Accepted 24 January 2003 / Received 25 November 2002}

\maketitle

\begin{abstract}
 There is growing observational  evidence that the irregular changes in
the  light  curves  of  certain   variable   stars  might  be  due  to
deterministic  chaos.  Supporting  these  conclusions,  several simple
models of  non-linear  oscillators  have been  shown to be  capable of
reproducing  the  observed   complex   behaviour.  In  this  work,  we
introduce a  non-linear,  non-adiabatic  one-zone  model  intended  to
reveal the factors leading to irregular luminosity  variations in some
pulsating  stars.  We have  studied and  characterized  the  dynamical
behaviour of the oscillator as the input  parameters  are varied.  The
parametric study implied values corresponding to stellar models in the
family of Long Period Variables and in particular of Mira-type  stars.
We draw  the  attention  on  certain  solutions  that  reproduce  with
reasonable  accuracy the  observed  behaviour  of some  peculiar  Mira
variables.
\end{abstract}

\begin{keywords}
stars:  AGB  and   post-AGB   ---  stars:  oscillations   ---   stars:
variables --- stars:  Miras
\end{keywords}

\section{Introduction}

The  region in  the  Hertzsprung--Russell diagram  which provides  the
majority  and, probably,  the  most interesting  classes of  pulsating
stars is the  Asymptotic Giant Branch (AGB).  As the  stars of low and
intermediate mass  (from say $\sim  1$ to $11\,M_\odot$)  evolve along
the  AGB phase,  they experience  recurrent thermal  instabilities and
substantial mass  loss.  Depending on  the phase in the  thermal pulse
cycle, they  may spend intermittent time intervals  as pulsating stars
(Groenewegen  \& Jong  1994).  As  a consequence,  they  eject freshly
synthesized material  into the  interstellar medium.  Thus,  they also
play a crucial role for our understanding of the chemical evolution of
galaxies.  Last  but not least,  the corresponding stellar  models are
basic tools to  study how planetary nebulae and  their central objects
form  --- see Habbing  (1996) and  Willson (2000)  for reviews  on the
subject.

\begin{figure*}
\vspace{3cm} 
\includegraphics{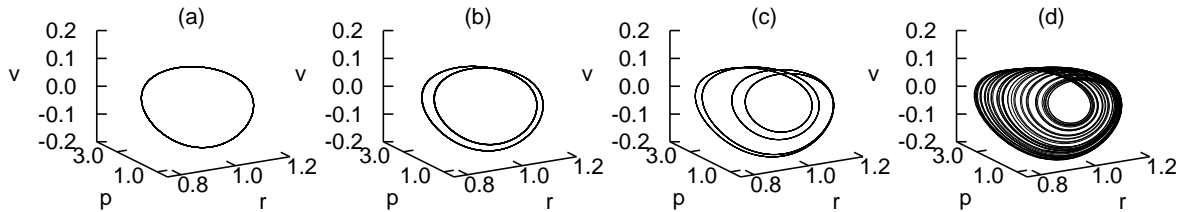}
\caption[]{Period-doubling  route to chaos  represented  in the  space
	  $(r,v,p)$.  In all panels, $\alpha=0.1$,  $\omega=20.1$, and
	  $a=20$ have been adopted.  {\sl (a)}  $\xi=0.08$:  Period-1;
	  {\sl  (b)}  $\xi=0.09$:  Period-2;  {\sl  (c)}  $\xi=0.108$:
	  Period:4; {\sl (d)} $\xi=0.12$:  Chaos.}
\label{fig01}
\vspace{-0.2cm}
\end{figure*}

A substantial  number of pulsating  stars in the Galaxy show irregular
behaviour and have the characteristics of AGB stars (Wood 2003).  This
makes them an interesting  research target for  theoretical  modeling.
The work done so far on non-linear  stellar  pulsations can be grouped
into two  different  categories:  the  full  numerical  hydrodynamical
approach  and the  more  straightforward  approach  of the  theory  of
dynamical  systems.  The  former  is  directly  based on an  extensive
knowledge of the physical processes in the star --- equation of state,
opacities,  and many more --- and  requires  the  implementation  of a
hydrodynamical  code coupled with a detailed treatment of the transfer
of  radiation.  Hence,  this  approach  definitely  provides  the most
detailed and accurate description of pulsations.  However, within this
framework it is sometimes difficult to interpret the origin and shapes
of the resulting  light curves as the stellar  parameters  are varied.
On its hand, the second approach is complementary to the former in the
sense  that it gives a  qualitative  framework  in which  the  general
features of the pulsations are easily  understood  and,  consequently,
allows to develop intuitive  explanations in terms of a few very basic
and relatively simple physical processes (Buchler 1993).

The   hydrodynamic   simulations  of  pulsating  stars   succeeded  in
reproducing  the  phenomenon of  period-doubling  (Buchler \& K\'ovacs
1987; K\'ovacs \& Buchler 1988; Aikawa 1990) and tangent  bifurcations
(Buchler,  Goupil \& K\'ovacs  1987;  Aikawa  1987) found in the light
curves of some specific stars by changing the surface  temperature  or
surface  gravity.  In  particular,  it has been  found  that  there is
strong   evidence  of  underlying   low-dimensional   chaos  from  the
non-linear  analysis of the light curves of the irregular stars R~Sct,
AC~Her, and SX~Her (Buchler,  Koll\'{a}th \& Cadmus 2001).  This paved
the road to construct simple models within the framework of the theory
of dynamical systems which turned out to be capable of reproducing the
complex  behaviour.  Within this approach  perhaps the simplest models
of stellar  pulsations  are the  so-called  one-zone  models.  In this
family of models the star is treated as a rigid core  surrounded  by a
homogeneous  gas shell.  The one-zone  models are not intended to be a
substitute  for  fully   hydrodynamical   and  sophisticated   models.
Instead, they merely seek a global and qualitative explanation of what
process might be at the root of the chaotic behaviour.  In the present
paper, we propose a one-zone model of non-linear non-adiabatic stellar
oscillations.  We present the parametric  study of the system and draw
the  attention on a particular  set of numerical  solutions  which may
have  implications  in the  study of the  variability  of Long  Period
Variables (LPVs).

\begin{figure*}
\vspace{6.5cm} 
\includegraphics{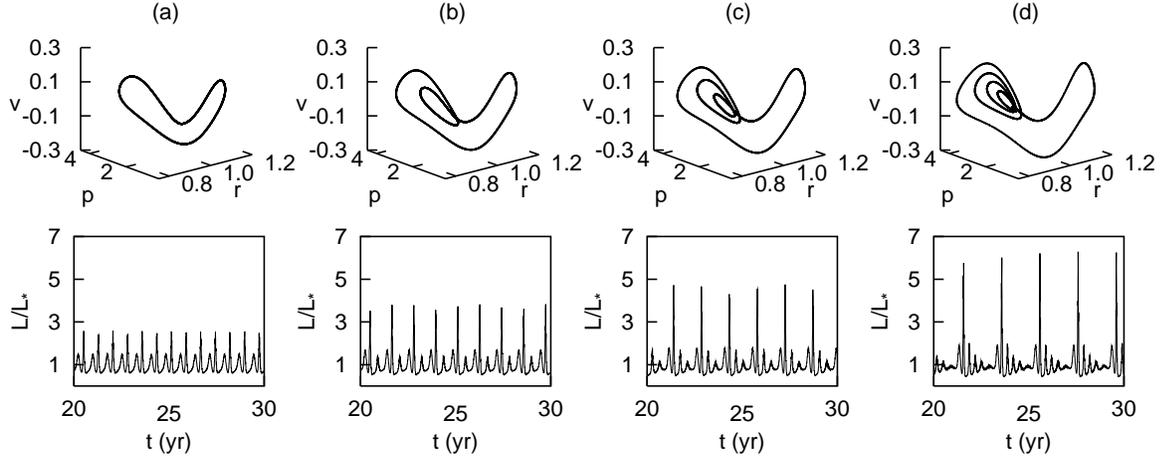}
\caption[]{The birth of a knot-like structure for increasing values of
	   $Q$.  {\sl Top panels:}  Stroboscopic sampling of the orbit
	   ($r,v,p$) with the characteristic frequency $\omega$:  {\sl
	   (a)} $Q=1.02$, {\sl (b)}  $Q=1.04$,  {\sl (c)} $Q=1.08$ and
	   {\sl (d)} $Q=1.16$.  {\sl Bottom panels:}  The light curves
	   corresponding to the cases shown in the upper panels.}
\label{fig02}
\vspace{-0.2cm}
\end{figure*}

\section{The one-zone model}

Baker et al. (1966) were the  first to introduce the one-zone model as
tool to study the  nonlinear behaviour of stellar pulsations.  Buchler
\& Regev (1982)  and Auvergne \& Baglin (1985)  later studied one-zone
models  under the assumption  that the  nonlinearity of  the adiabatic
coefficient $\Gamma_1$  is the main trigger  for nonlinear pulsations,
whereas Tanaka \& Takeuti  (1988) pointed out that dynamical stability
might be necessary  for realistic models of pulsating  stars, which is
the approach followed here.

The one-zone model approach is  especially suited for the AGB stars as
the density difference  between the central core and  the outer layers
is so  large that these two  regions can be  considered as effectively
decoupled.  Following Icke et al.  (1992) and Munteanu et al.  (2002),
we consider the stellar pulsation  to be simulated by a variable inner
boundary located well beneath the photosphere and moving with constant
frequency (piston approximation).  This sinusoidal driving consists of
pressure waves  originating in the interior and  propagating through a
transition  zone  until  they  dissipate  in  the  outer  layers.   We
denominate  the driving  oscillator  ``the interior''  and the  driven
oscillator  ``the mantle''.  While  the radius  of the  former region,
$R_0$, can be  estimated as the radius of  the hydrogen burning shell,
the radius of the latter, $R$  is determined by the dissipation of the
aforementioned pressure waves.
 
The  variation  of  the  interior  radius,  $R_{\rm  c}$,  around  the
equilibrium value, $R_0$, is given by $R_{\rm  c}=R_0+\alpha R_0 ~\sin
~\omega_{\rm  c}  \tau$,  where  $\alpha$  and  $\omega_{\rm  c}$ are,
respectively,  the  fractional  amplitude  and  the  frequency  of the
driving.  As in Stellingwerf (1972), we use the equation of motion and
the energy  equation  without  energy  sources  and in absence  of any
driving  force in order to  determine  the final  equation  of  motion
describing  the dynamics of the mantle.  In the present  work, we will
study the particular  case in which the  luminosity at the base of the
mantle  equals  the  (constant)  equilibrium  luminosity  of the star,
$L_{\star}$.  For convenience, we use as variables the non-dimensional
radius $r \equiv R /R_\star$,  pressure $p \equiv  P/P_\star$ and time
$t \equiv  \omega_{\rm  m}\tau$, where the stellar radius and pressure
were normalized to their equilibrium values, while

\begin{equation}
\omega_{\rm m} \equiv \sqrt{\frac{GM}{R_\star^3}}
\end{equation}

\noindent is the characteristic frequency of the star.

Considering   that  the   additional   perturbative  acceleration   is
proportional   to  the  driving   acceleration  with   a  transmission
coefficient $Q$ as in Icke  et al.  (1992) and following the reasoning
of Saitou, Takeuti \& Tanaka (1989) for the energy equation, we obtain
the following equation of motion

\begin{eqnarray}
\frac{dr}{dt}&=&v   \nonumber  \\ 
 \frac{dv}{dt}&=&   p\,r^2  -r^{-2}-\nonumber  \\  
\lefteqn{~~~~~-Q\alpha\omega^{4/3}\sin(\omega r-\omega t-\alpha\omega^{1/3}
\sin \omega t)}
\label{eq:final}\\
\frac{dp}{dt}&=& -3\Gamma_1 r^{-1}vp-\xi r^{-3} (r^\beta p^\delta -1),
\nonumber
\end{eqnarray}

\noindent where

\begin{equation}
\beta = a(r^3p-1.2) +21.6 ~~,~~\delta = 3.6 r^3 p (r^3 p -0.2)
\label{eq:delta}
\end{equation}

\noindent are the coefficients introduced in Saitou et al. (1989) as a
saturation effect of the $\kappa$-mechanism with $a$ being the control
parameter.  The characteristic frequency of the perturbation, $\omega$
results  from the  use of  the dimensionless  time unit  and  from the
assumption that $R_0$ encompasses  almost the entire stellar mass.  It
is defined as

\begin{equation}
\omega  \equiv \frac{\omega_{\rm  c}}{\omega_{\rm m}}  =  \left( \frac
{R_0}{R_ \star}\right) ^{-3/2}.
\label{eq:omega}
\end{equation}

In  its  simplest  form,  the  study  of  stellar  pulsations  can  be
considered   as  a   thermo-mechanical,  coupled   oscillator  problem
(Gautschy \&  Glatzel 1990).   The coupling constant  is given  by the
ratio of the  dynamical to the thermal time scale  in the outer layers
of the  star.  Whenever  the thermal time  scale ($\tau_{\rm  th} \sim
4\pi r^2 \rho  \Delta r c_{\rm V} T/L_{\rm r}$) of  an outer region of
the star of  radial extension $\Delta r$ happens  to become comparable
to the sound-traveling time  through that region ($\tau_{\rm dyn} \sim
\Delta  r/c_{\rm  s}$),  non-adiabatic  effects  are  relevant.   This
implies that there is an  efficient exchange of mechanical and thermal
energies in that region.  The ratio of the time scales is much smaller
than one throughout most of the envelope and is close to unity only in
the outermost regions.  Significant non-adiabatic effects are relevant
for helium stars,  very massive AGB stars, some  post-AGB stars and in
the ionization zones of hot stars (Stellingwerf 1986; Gautschy \& Saio
1995).   In  the  context  of  our  system,  the  parameter  $\xi$  in
Eq.(\ref{eq:final}) is a measure  of the non-adiabaticity and is given
by the  ratio of the dynamical  timescale to the  thermal timescale of
the shell

\begin{equation}
\xi ~=~ \frac{L_\star}{\omega_m c_{\rm V} m T}.
\end{equation}

The  system  of  Eq.(\ref{eq:final})  constitutes  the  final  set  of
relations for the unknowns $r$, $v$ and $p$.  The parameters that must
be specified  are $Q$, $\alpha$, $\omega$, $\Gamma_1$,  $a$ and $\xi$.
We consider an ideal gas with an adiabatic coefficient $\Gamma_1=5/3$.
The parameter fixing  the evolutionary status of the  star is $\omega$
as it  provides a  measure of  the contrast between  the core  and the
envelope.  For AGB stars, $R_0/R_\star$ is of the order of 15\%.  This
led Icke et al.  (1992) to  consider a value of $\omega=20.1$ which is
the  value  adopted  here.   We  have  performed  a  parametric  study
involving the  parameters of the perturbation, $Q$  and $\alpha$.  The
results of the numerical  integrations show that the pair $(Q,\alpha)$
--- and not  the individual  specific values of  $Q$ and  $\alpha$ ---
determines the dynamics of the  system.  In other words, given a fixed
value of $\alpha$,  a certain range of values of $Q$  can be found for
which the same peculiar dynamics discussed below develops.  Therefore,
we present here only the  case of small internal perturbation ($\alpha
\approx 3-4\%$)  and amplified  transmission through the  envelope $(Q
>1)$.   It is worth  noticing that  for higher  (lower) values  of the
parameter  $\alpha$,  the  same  behaviour  is  encountered  if  lower
(higher) values of the coupling coefficient $Q$ are chosen.

\section{Numerical results}

\begin{figure*}
\begin{center}
\vspace{8.5cm}  
\includegraphics{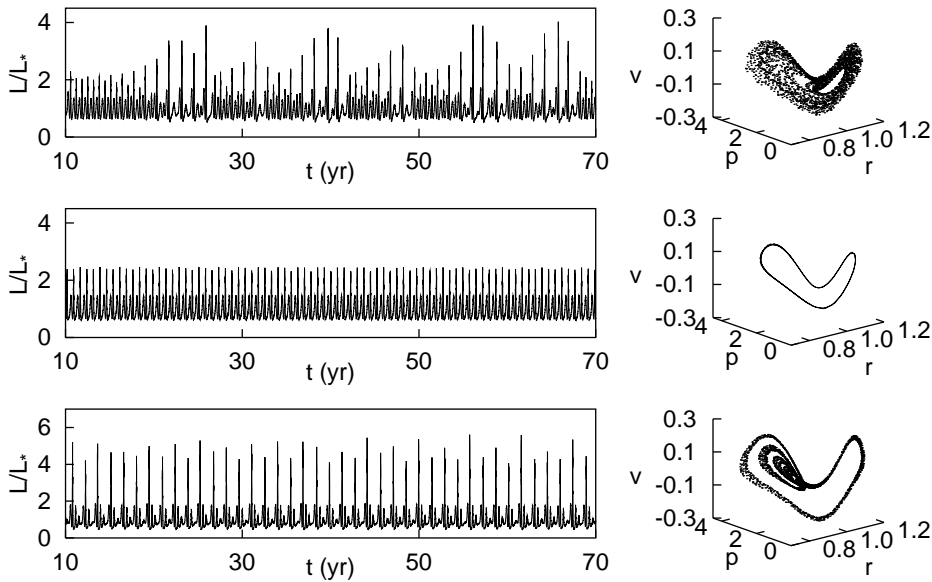}
\caption[]{Comparison  between  regular and  irregular  dynamics.  The
	   light curves and the  corresponding  stroboscopic  maps for
	   the  cases  $\alpha=0.037$,  $\omega=20.1$  and  {\sl  (top
	   panels)}  $Q=0.8$, {\sl (middle  panels)}  $Q=1.0$ and {\sl
	   (bottom panels)} $Q=1.048$.}
\label{fig03}
\end{center}
\vspace{-0.7cm}
\end{figure*}

\subsection{Zero perturbation}

The system  studied by Saitou et al.  (1989) can be obtained  from the
system  given in  (\ref{eq:final})  by  considering  the  case of zero
perturbation   ($Q=0$).  In  spite  of  the  inexistence  of  internal
perturbation, this system is the prototype of non-linear  self-excited
oscillators  due the  $\kappa$-mechanism  provided  that $\xi \neq 0$.
The behaviour of any  dynamical  system  $\mbox{\boldmath$\dot{x}$}  =
\mbox{\boldmath$F$}  (\mbox{\boldmath$x$})$, with $\mbox{\boldmath$x$}
= (x_1,x_2,...,x_n)$ and  $\mbox{\boldmath$F$} = (F_1,F_2,...,F_n)$ is
critically  determined  by its  fixed  points  $\mbox{\boldmath$x$}_0$
given   by   $\mbox{\boldmath$F$}(\mbox{\boldmath$x$}_0)   =  0$.  The
associated     eigenvalues    of    the    Jacobian     matrix    $(J_
{\mbox{\boldmath$x$}_0})_{ij}=     (\partial     F_i/\partial    x_j)_
{\mbox{\boldmath$x$}_0}$  determine  the nature of these fixed points.
For instance, for a fixed point to be stable, it is required  that all
eigenvalues  have  negative  real  parts.  In our case, the system has
three fixed  points:  a trivial one,  $(r_0,v_0,p_0)=  (1,0,1)$,  with
mainly    adiabatic    origin    and   two   other    fixed    points,
$(r_0^+,v_0^+,p_0^+)  \approx (0.68,0,4.75)$ and  $(r_0^-,v_0^-,p_0^-)
\approx (8.8558,0,0002)$,  entirely due to non-adiabatic effects (that
is, they exist only for $\xi \neq 0$).  For initial  conditions  close
to the trivial  fixed  point, the  period-doubling  route to chaos was
obtained  by  Saitou  et  al.  (1989)  by   decreasing   the   surface
temperature,  more  precisely by varying the control  parameter $a$ in
the  range  $a  \in  [14,20]$,  while  $\xi$  was  kept  constant.  No
investigation  of an equivalent  effect  produced by varying $\xi$ was
carried on.  Concerning the dynamics near the trivial fixed point, our
analysis of the  dependence  of the  Jacobian  matrix on $a$ and $\xi$
reveals  that for  $a\la  36$,  either  the  increase  of $\xi$ or the
decrease of $a$ leads to the same effect on the eigenvalues consisting
in chaotic behaviour through the increase of pulsational  instability.
In order to prove  the  importance  of $\xi$  in the  dynamics  of the
system and to  complete  the study of Saitou et al.  (1989), in Figure
\ref{fig01}  we  present  a  period-doubling  route to chaos  with the
increase of the parameter $\xi$ as it appears in the space  $(r,v,p)$.
In order to ease the exploration of the dynamic details, we use mainly
the Poincar\'{e}  map.  As we deal with a periodically  driven system,
the  Poincar\'{e}  map reduces to a stroboscopic  sampling of the $r$,
$v$, and $p$ values at multiples of $T=2\pi/\omega$.

\subsection{Non-zero perturbation}

We present  now the properties  of the non-linear oscillator  that are
due to the presence of the time-dependent perturbation.  We fix $a=20$
as  corresponding  to  the   regular  pulsation  found  by  Saitou  et
al. (1989)  and, moreover, we choose  $\xi =0.06$ in  order to compare
with their work.  Additionally  we fix $\omega=20.1$ and $\alpha\simeq
0.037$.   Our study  considers  the coupling  coefficient  $Q$ as  the
primary control parameter for  the strength of the perturbation, while
$\omega$  and $\alpha$  are  kept constant  at  the values  previously
mentioned.  In  the top  panels of Figure  \ref{fig02} we  present the
stroboscopic  map of  the system  for  increasing values  of $Q$.   We
notice  the  successive  creation  of  loops,  finally  leading  to  a
knot-like structure for strong  perturbation ($Q=1.16$).  For the sake
of comparison with real astronomical  data, we also plot in the bottom
panels of  Figure \ref{fig02} the  light curves for  the corresponding
cases.  The luminosity $L$ is obtained from the condition of radiative
transfer together with the perfect gas law (Stellingwerf 1972):

\begin{equation}
\frac{L}{L_\star} = r^\beta p^\delta ,
\end{equation}

\noindent where the normalizing  constant $L_\star$ is the equilibrium
stellar luminosity.  The temporal scale  is expressed in years and for
this task we have used  the stellar parameters associated to a typical
Mira of  $1 M_\odot$ as  they result from  the work of  Vassiliadis \&
Wood (1993).

The most noticeable feature of  the light curves of Figure \ref{fig02}
consists in a highly energetic  sporadic burst followed by a series of
smaller peaks.   The number  of small peaks  which appear  between the
major ones depends on $Q$.  Moreover, the creation of every new inward
loop in the stroboscopic map is  equivalent to the appearance of a new
small peak in the light curve.   We also found that each change in the
number  of loops  is accompanied  by a  chaotic regime.  That  is, the
transition from  a regular  regime to a  chaotic one occurs  through a
sequence   of   period-doubling   bifurcations   similar   to   Figure
\ref{fig01}. 

For  a  clear  exemplification  of  the  chaotic  regimes,  in  Figure
\ref{fig03} we  show the light curves and  the associated stroboscopic
maps for  a case of  regular dynamics and  for two cases  of different
degrees of irregularity.  The central panels correspond to the case of
regular dynamics prior  to the development of the  successive loops in
the stroboscopic map.  For a slightly lower value of the parameter $Q$
(top  panels), a  completely  irregular light  curve  results and  the
stroboscopic map clearly  shows it.  Finally, in the  lower panels, we
illustrate  the  dynamics  of  the  system  when  $Q$  takes  a  value
corresponding  to the  accumulation of  a sequence  of period-doubling
bifurcations.

\begin{figure*}
\begin{center}
\vspace{6.5cm} 
\includegraphics{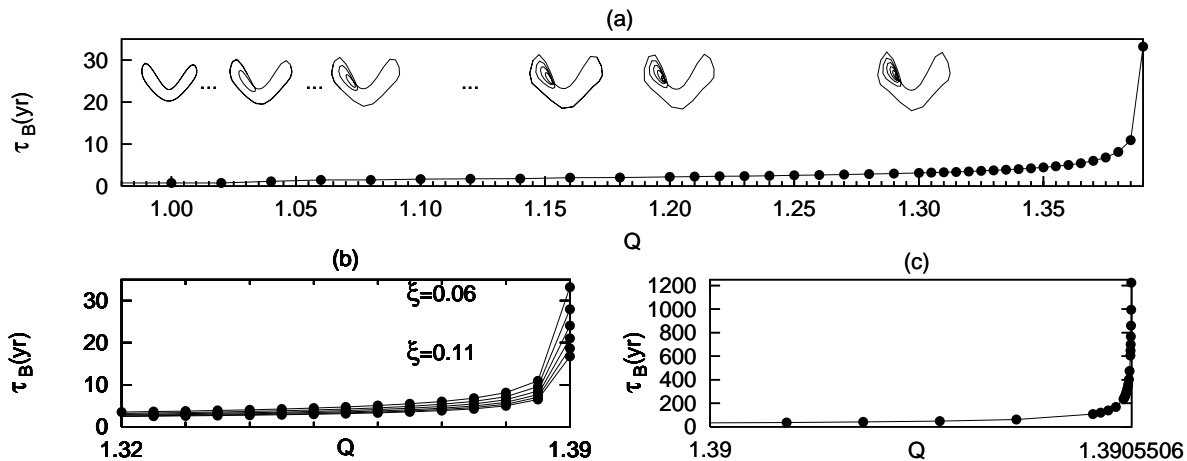}
\caption[]{Time interval between bursting oscillations.  In all cases,
	   $\alpha=0.037$,   $\omega=20.1$,   $a=20$.  {\sl(a)}   Time
	   between   major  peaks  as  a  function  of  the   strength
	   coefficient  $Q$ for the  case  of  $\xi=0.06$;  {\sl  (b)}
	   Variation of the time between  major peaks as a function of
	   $Q$ and for  different  values of $\xi$  from 0.06 to 0.11;
	   {\sl(c)}  The  extension  to higher  values of $Q$ leads to
	   larger time intervals between major peaks for $\xi=0.06$.}
\label{fig04}
\end{center}
\vspace{-0.8cm}
\end{figure*}

Another important  feature of the dynamics  is the fact  that the time
interval  $\tau_{\rm  B}$  between  major bursts  increases  with  the
strength of the  perturbation, as it can be seen  in the bottom panels
of Figure \ref{fig02}.  This  is illustrated in Figure \ref{fig04}{\sl
a}, where  the time  interval between the  successive major  bursts is
shown  as a function  of the  coupling coefficient  $Q$.  As  a visual
guide we  also plot  the shape  of the stroboscopic  map at  the fixed
values of  $Q$ where an additional  inner loop appears.  As  it can be
seen in  this panel  and in panel  \ref{fig04}{\sl b},  $\tau_{\rm B}$
significantly increases  for $Q\in[1.32,1.39]$ whereas for  $Q > 1.39$
(Figure \ref{fig04}{\sl  c}) the separation  between successive bursts
increases drammatically.   Also, in Figure \ref{fig04}{\sl  b} we show
that  the above  mentioned  increase  of $\tau_{\rm  B}$  can also  be
obtained by  decreasing the non-adiabaticity  of the system  (that is,
decreasing  $\xi$).  Nevertheless  the main  parameter for  tuning the
time interval between major bursts turns  out to be $Q$.  Note as well
that these  major bursts favor  mass loss at exceptionally  high rates
and, moreover, the time intervals between them are long.  Hence, it is
tantalizing to  directly connect them with  the periodicities observed
in  the  circumstellar  shells  that  can be  found  surrounding  some
planetary nebulae (Van Horn et al.  2002).

\subsection{Mathematical interpretation of the results}

In order to validate the numerical results discussed above, we briefly
present the  mathematical  characteristics  of our system.  Given that
our  system  is  non-autonomous   ---  that  is,  the  Hamiltonian  is
explicitly  time-dependent  --- the typical methods of analysis of the
theory dynamic  systems cannot be used.  To overcome this drawback, we
used an averaging  method  (Sanders \& Verhulst 1985) to transform our
system into an autonomous  one.  The high value of the  characteristic
frequency  ($\omega \gg 1$) assures us that this method is  applicable
to our  case.  In  the  time-averaged  framework,  the  time-dependent
perturbation  from  Eq.  (\ref{eq:final})  becomes  $F(r) = -Q  \alpha
\omega^{4/3}~  A~ \sin  \omega r$, where  $A=-0.04993$  is a  constant
evaluated numerically.  The fixed points of the averaged system in the
case of nonzero perturbation must satisfy the conditions:

\begin{eqnarray}
&&p_0= r_0^{-4}+Q \alpha \omega ^{4/3} A ~ r_0^{-2}~ \sin ~\omega r_0
\label {eq:p0condition}\\ 
&&r_0^\beta p_0^\delta -1=0~.
\end{eqnarray} 

\noindent or, equivalently,

\begin{equation}
G(r)  \equiv r^\beta  (r^{-4}+Q \alpha  \omega ^{4/3}  r^{-2}  A~ \sin
~\omega r)^\delta-1=0.
\label {eq:PF}
\end{equation}

\noindent   The  roots  of   Eq.   (\ref{eq:PF})   are  to   be  found
numerically. The new  fixed points of interest are  $r_0 \approx 0.65$
and $r_0 \approx 1.03$ and  represent only slight displacements of the
fixed  points  mentioned before  in  the  case  of zero  perturbation.
Moreover,  their associated  eigenvalues  maintain the  form from  the
previous   case  (unstable   fixed  points   of   saddle-focus  type).
Increasing the  parameter $Q$, a new  fixed point is  created at about
$Q_0  \approx 1.3855$  at $r\approx  0.86$.  This  new fixed  point is
stable  as the associated  eigenvalues have  all negative  real parts.
With  the  creation  of   this  fixed  point,  the  looping  behaviour
dissappears. For higher values of  $Q$, the fixed point is replaced by
two unstable  fixed points again  of saddle-focus type.   The distance
between  them  increases with  $(Q-Q_0)^{1/2}$.   Note, however,  that
these  very high  values of  $Q$ result  in transmissions  through the
mantle which are not very  realistic and, moreover, the time intervals
between successive bursts for larger values of $Q$ become very large.

\subsection{The role of ~$\omega$}

\begin{figure*}
\begin{center}
\vspace{8.3cm} 
\includegraphics{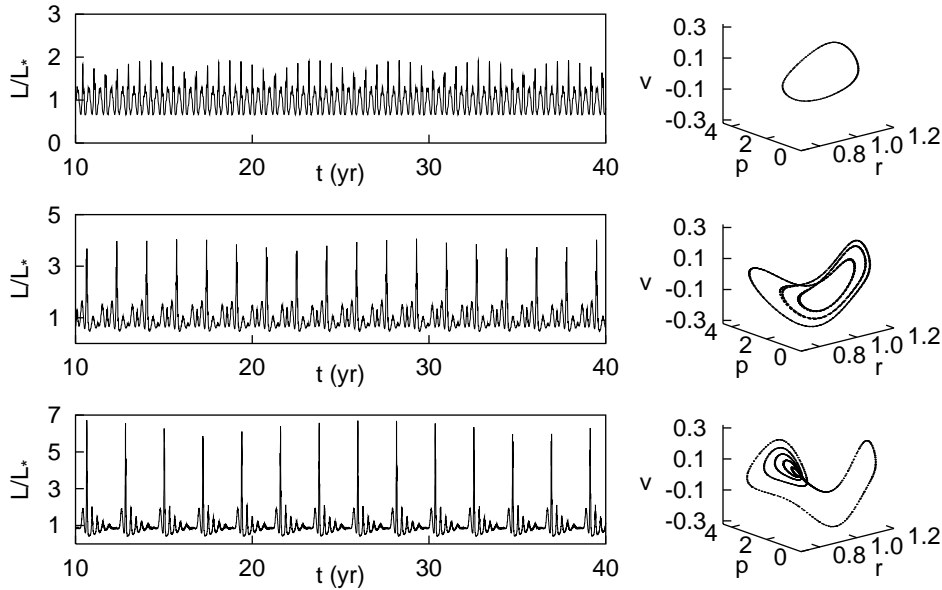}
\caption[]{The   role  of   $\omega$.  The   light   curves   and  the
	   corresponding    stroboscopic    maps    for   the    cases
	   $\alpha=0.037$,  $Q=1.2$ and {\sl (top panels)} $\omega=10$
	   ($M=5 M_\odot$), {\sl (middle  panels)}  $\omega=15$  ($M=3
	   M_\odot$)  and {\sl (bottom  panels)}  $\omega=20.1$  ($M=1
	   M_\odot$).}
\label{fig05}
\end{center}
\vspace{-0.7cm}
\end{figure*}

Before interpreting the behaviour of the system in the context of real
data of stellar variability, we  consider worth of interest to explore
the role of the parameter $\omega$ and to justify the particular value
we have used throughout this work.  Icke et al.  (1992) concluded that
in the case of complete adiabaticity ($\xi=0$), a decrease of $\omega$
leads to stronger chaotic pulsations.   The values of $\omega$ used in
their work were equivalent to adopting stellar models in the family of
low-mass stars ($M  \leq 5-8 M_\odot$) reaching the  AGB.  In a recent
publication (Munteanu  et al., 2002), we extended  their conclusion to
intermediate-mass stars ($8  M_\odot \leq M \leq 11  M_\odot$) also in
the AGB phase, more precisely to  values of $\omega$ around 3.  In the
previous sections, we  have shown that in the  case of $\omega=20.1$ a
peculiar behaviour is born from the interplay between non-adiabaticity
and internal  perturbation.  Our analysis of  the system corresponding
to the parametric  interval $5 \leq \omega \leq  25$ has revealed that
such a behaviour is found for  values of $\omega$ close to 20, that is
for low-mass  stars.  Mathematically, we attribute  this fact entirely
to the creation of new  fixed points mentioned in the previous section
which critically  alter the  dynamics of the  system.  They  exist for
values of  $\omega$ higher than  about 18.  For values  slightly lower
than 18, the dynamics  resembles the one encountered for $\omega=20.1$
(middle panels  of Figure  \ref{fig05}), but it  does not  present the
successive creation of new loops.   Instead, the change of the control
parameter $Q$ leads  to a mixture of chaotic  regimes and uncorrelated
creation and dissapearance of new loops.  For completeness, we present
in Figure \ref{fig05}  the light curves and the  stroboscopic maps for
three values of $\omega$.  For  each case, the temporal scaling factor
was computed according to the work of Vassiliadis \& Wood (1993) which
provides  a complete  set of  stellar  parameters for  AGB stars  with
initial masses  in the range $0.89  \leq M/M_\odot \leq  5.0$.  Due to
the richness of  the dynamics in the case of  $\omega=20.1$, we use it
for a comparison with observations.

\section{Comparison with observations}

The  values  of  the   parameters   used   throughout   the  numerical
integrations were intended to locate the stellar models we are dealing
with in the family of Long Period Variables (LPVs) and,  specifically,
in the families of semiregular  and Mira  variables.  Perhaps the most
important  of the Mira  stars is  $o$~Ceti.  Its light  curve  shows a
peculiar variability consisting in an exceptional peak occurring every
two, three or five ``cycles''  (Barth\'es \& Mattei 1997).  Our simple
model naturally  recovers this behaviour by tuning the strength of the
perturbation  or the  coupling  coefficient.  Moreover,  we have shown
that  within  our  simple  model  large  peaks in the light  curve are
associated to large inter-pulse intervals.

\begin{figure*}
\vspace{-0.3cm}
\hspace{-0.5cm}
\epsfxsize=1.3\textwidth
\includegraphics[clip,width=7.2cm,angle=-90]{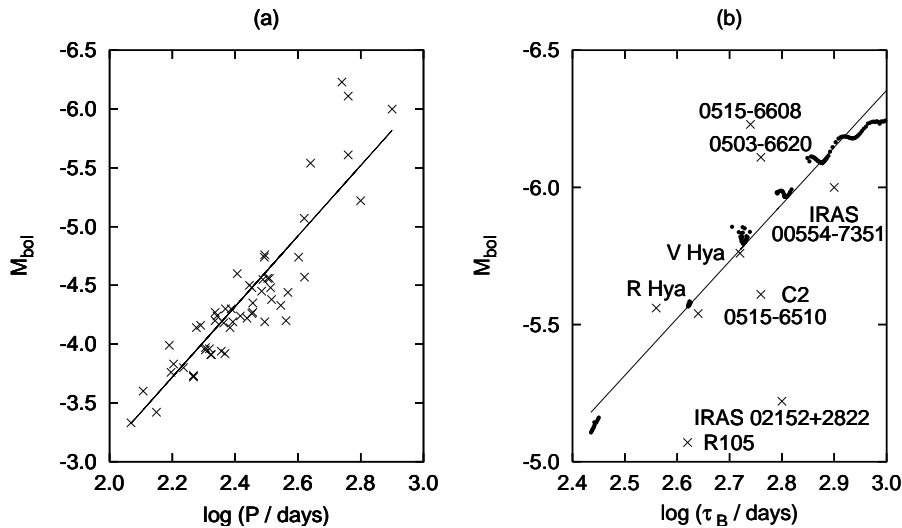}
\caption[]{Period-Luminosity  relation  for Mira  stars.  {\sl (a)} PL
	   relation  for Mira  stars in the LMC and the  data on which
	   the  relationship  is based (Feast 1989).  {\sl (b)} The PL
	   relation  using the data  furnished  by our  model  (filled
	   circles):  the equivalent  $M_{\rm bol}$ of the major peaks
	   and the time  interval  $\tau_{\rm  B}$ between  them.  The
	   values  for the  parameters  used here are  $\alpha=0.037$,
	   $\xi=0.06$ and $Q \in  [1,1.26]$.  The crosses  represent a
	   sample of  over-luminous  Mira  stars in the LMC.  See text
	   for additional details.}
\label{fig06}
\vspace{-0.3cm}
\end{figure*}

The Mira stars belonging to the Large Magellanic Cloud  constitute the
best sample of Miras concerning both  periodicities  and luminosities.
In Figure  \ref{fig06}{\sl a} we show the observational  data for some
Mira  stars  in  the  LMC  (Feast  1989)  and  the  best  fit  to  the
observational  data.  Among them, the ones having periods  longer than
about 400 days clearly appear to be over-luminous  with respect to the
period-luminosity  relationship  found for Miras with relatively short
periods  (Zijlstra et al.  1996;  Bedding et al.  1998).  With this in
mind we  have  obtained  a  period-luminosity  relationship  from  our
theoretical models using the time interval $\tau_{\rm B}$ and averaged
values of the  major  peaks in  luminosity.  More  precisely,  we have
varied the parameter  $Q$ in the range  $[1,1.26]$  resulting in light
curves whose major peaks have periodicities (in days) within the range
$2.4\la\log\tau_{\rm   B}\la   3.0$.  The   corresponding   bolometric
magnitudes were computed using a reference  value for the  equilibrium
stellar  luminosity of  $\log(L_\star/L_\odot)  = 3.5$ (Vassiliadis \&
Wood  1993)  which  is  typical  for  Mira   stars.  Our   theoretical
period-luminosity  relationship is shown in Figure \ref{fig06}{\sl b}.
For the sake of comparison  we also show in this panel all those Miras
with  $\log P> 2.4$ and  $M_{\rm  bol} < -5.0$,  that is all the stars
which are found to be over-luminous in Figure  \ref{fig06}{\sl a}.  We
have also included two other interesting  Miras:  R~Hya and V~Hya.  As
it can be seen our  theoretical  period-luminosity  relationship  fits
very well the  observational  data.  Moreover,  note as well  that the
observations   tend  to   cluster   around   fixed   regions   of  the
period-luminosity relationship.  These regions are coincident with the
regions  where we find  regular  oscillations  with a fixed  number of
loops in the  stroboscopic  map and its  sequence  of  period-doubling
cascade leading to chaos.  The gaps between the  theoretical  data are
the  consequence of the drastic change in the  characteristics  of our
light  curves when a new loop is  created,  and  correspond  to a very
small change of the  coupling  coefficient.  Hence,  according  to our
analysis we should not find almost any star in these regions, which is
exactly what it is found.

Another  peculiarity  of  Mira  stars   is  that  some  of  them  show
alternating deep  and shallow minima, giving the  appearance of double
maxima.  Some examples are R~Cen,  R~Nor, U~CMi, RZ~Cyg and RU~Cyg ---
see Hawkins (2001) and references therein.  Among these, R~Cen has the
most persistent and stable double  maxima in the light curve while for
the rest of  the cases the second maximum is often  weak and the light
curve  sometimes reverts  to that  of a  normal Mira.   Our simplistic
model provides  such light  curves for small  values of  the parameter
$Q$.   As  for  the   general  ``chaotic  connection''  for  the  Mira
variables,  the  first  (and  unique)  case  of  evidence  of  chaotic
pulsation  in a  Mira star  (R~Cyg) comes  from the  study by  Kiss \&
Szatm\'{a}ry  (2002).    They  associate  the   long  sub-segments  of
alternating  maxima in  R~Cyg to  a period-doubling  event, supporting
therefore the  well-known scenario of period-doubling  to chaos, which
we also find in our model.  Buchler et al.  (2001) present an overview
of  observational examples  of chaotic  behaviour in  some semiregular
variables  (SX~Her, R~UMi, RS~Cyg,  and V~CVn).   They argue  that AGB
stars are  prone to chaotic pulsations  due to the  fact that relative
growth rates of the lowest frequency  modes are of the order of unity.
Higher   relative   growth  rates   are   a   consequence  of   higher
luminosity--mass ratios, that is more non-adiabatic stars.  Hence, not
only  semiregular  variables,  but  also  Mira stars  should  also  be
candidates  for chaotic pulsators.   In spite  of its  simplicity, our
toy-model  presents chaotic  pulsations for  certain intervals  of the
parameters characterizing  the strength  of the internal  driving and,
thus, could provide some support  to the conjecture that the evolution
of semiregulars and of Mira stars is strongly connected.

\section{Conclusions and caveats}

We have  introduced a weakly  non-adiabatic  one-zone  model driven by
sinusoidal   pressure  waves   intended  to  reproduce  the  irregular
pulsations  of  Mira-like  variables  for which  other  simple  models
already exist (Buchler \& Regev 1982; Auvergne \& Baglin 1985; Reid \&
Goldston  2002).  Our model  extends  the works of Icke et al.  (1992)
and  Saitou  et  al.  (1989).  In  particular,   Icke  et  al.  (1992)
proposed an  adiabatic  model  driven by  pressure  waves (the  piston
approximation)   whereas  Saitou  et  al.  (1989)   studied  a  simple
non-adiabatic  model  without  driving  (the  self-excited   pulsation
model).  Our  approach  is  justified  by the large  density  contrast
between  the  interior  and the outer  layers.  One-zone  models  have
limitations  since some Mira stars are  suspected to have at least two
frequencies (Mantegazza 1996) which may vary independently, suggesting
that more than one mode is  involved.  Clearly,  one-zone  models  are
unable to reproduce  this  behaviour.  Nevertheless,  the  majority of
Miras do not show this behaviour.  Another interesting  approach would
have been to use the modal coupling.  However,  although this approach
has been used to model  the  pulsation  of Miras  (Buchler  \&  Goupil
1988), it is more  appropriate  for classical  Cepheids,  RR~Lyrae and
W~Vir stars --- see Buchler et al.  (1993) and references therein.  It
is also worth  noting  here that the  piston  approximation  --- first
introduced  by Bowen (1988) --- used in this paper has been used since
then by several  authors  (Fleischer et al.  1995;  H\"{o}fner  et al.
2003)  even if it has never  been  deeply  scrutinized  for  validity.
Nevertheless  this  approximation  appears to correctly  reproduce the
velocities  and mass loss rates  typical  of AGB  stars.  Hence,  this
approximation   can   be   regarded   as  a   reasonable   first-order
approximation  of  the  dynamical  effect  of  the  pulsation  on  the
atmosphere.  To  summarize,  our model should be  considered  as a toy
model that qualitatively  reproduces the general features of the light
curves.

We  have thoroughly  explored the  interesting particularities  of the
system from both the astrophysical and from the mathematical points of
view.  We have found that  the degree of non-adiabaticity turns out to
be  a determining  factor in  the development  of  the period-doubling
route to  chaos.  As far as  the transition to chaos  is concerned, an
increase in  the degree of non-adiabaticity  plays the same  role as a
decrease of the  effective temperature in the model  studied by Saitou
et al.   (1989).  Moreover, due to the  time-dependent perturbation, a
knot-like structure  is created  in the phase  space when  varying the
transmission coefficient ($Q$) while  the strength of the perturbation
is kept fixed.  The  resulting periodic light curves are characterized
by a  repetitive pattern  consisting in a  major peak followed  by $n$
minor peaks,  where $n$ is given by  the number of loops  in the phase
space.  Our theoretical light curves are in qualitative agreement with
those of  several well-known Mira stars.  In  particular the prototype
of Mira stars, $o$~Ceti, shows this pattern of alternating major peaks
followed by several  minor peaks.  Furthermore, for a  given choice of
our input parameters the resulting light curves also resemble those of
some  peculiar Miras (R~Cen,  R~Nor, U~CMi,  RZ~Cyg, or  RU~Cyg) which
appear  to have  double maxima  due  to alternating  deep and  shallow
minima. We have found as  well that our dynamical system presents both
chaotic  regimes and  patterns  of periodicity.   The chaotic  regions
occur for ranges of $Q$ between those corresponding to the creation of
a new luminosity peak between major bursts.  We have also noticed that
for increasing strengths of the perturbation the time interval between
major bursts increases.  This interval increases with $Q$ as well.  At
high  values of  the strength  of  the perturbation  within the  range
yielding the  knot-like structure,  we obtained a  peculiar behaviour:
the creation of new loops stops  and the resulting light curves show a
time interval during which  the luminosity preceeding every major peak
remains constant.

We have also obtained a theoretical period-luminosity relationship and
compared  it with  the observational  data of  Miras in  the  LMC.  In
particular, we have focused on  those peculiar Miras with long periods
which are  known to be over-luminous  with respect to the  best fit of
Feast (1989).  We have found  that our model provides a reasonable fit
to the period-luminosity  relationship of these stars and  also to the
observed clustering of the  over-luminous stars around certain regions
in the  period-luminosity diagram.  The ultimate reason  for this fact
is closely related  to the creation of a new  loop in the stroboscopic
map, and, consequently, of a  new luminosity peak in the corresponding
time series. 

Finally,  we would  like to stress  that  although  our  simple  model
succeeds  in  furnishing  reasonable  comparisons  with  real  data, a
detailed  quantitative   interpretation  is  beyond  its  capabilities
because of its crude physical  assumptions.  In particular, next steps
towards improving the one-zone  approach could eventually  include the
treatment of convection which is supposed to play an important role in
the envelopes of these stars.  This  improvement  will affect also the
functional form of the luminosity  which was rather  simplified in the
present work.

\vspace{0.2cm}

\noindent {\sl Acknowledgements.}  This work has been supported by the
MCYT grant  AYA2000--1785, by the MCYT/DAAD grant  HA2000--0038 and by
the CIRIT grants 1995SGR-0602 and  2000ACES-00017.  We would also like
to  acknowledge  our  anonymous  referee for valuable  criticisms  and
suggestions.

\end{document}